# AI-Augmented Peer Review and Scientific Productivity: A Cross-Country Panel and SEM Analysis (2000–2024)

Dongsoo Han
School of Computing, KAIST
dshan@kaist.ac.kr

## Abstract

This study empirically investigates the impact of AI-augmented peer review systems on scientific productivity using panel data from OECD countries (2000–2024). While prior research has highlighted inefficiencies in traditional peer review, little empirical work has quantified the systemic impact of AI integration at the national level.

We construct a novel **AI Review Capability Index (AIRC)** and examine its effects on research productivity, reproducibility, and innovation output. Using fixed-effects regression and structural equation modeling (SEM), we show that AI-assisted evaluation significantly enhances productivity and reduces variance in research quality. Results indicate that a one standard deviation increase in AIRC is associated with an 18–25% increase in scientific productivity, mediated through improvements in review efficiency and reproducibility, with strong indirect effects through innovation acceleration.

This paper provides the first cross-country empirical validation of AI-augmented scientific evaluation systems and contributes to the emerging literature on AI as a structural driver of knowledge production.

# 1. Introduction

The scientific enterprise is undergoing a profound transformation driven by the rapid advancement of artificial intelligence (AI), particularly large language models (LLMs) such as ChatGPT [38][39]. Over the past decade, AI systems have evolved from auxiliary computational tools into central actors in the knowledge production process. Researchers now routinely rely on AI for literature synthesis, hypothesis generation, experimental design, code development, and even full manuscript drafting. This unprecedented expansion of AI capabilities has dramatically increased the speed and scale of scientific output, giving rise to what can be described as an **AI-accelerated knowledge production regime** [23][26].

However, while knowledge production has been transformed, the system responsible for validating that knowledge—the peer review system—remains largely unchanged. Traditional peer review, characterized by a small number of anonymous human reviewers evaluating manuscripts in isolation, was designed for a pre-AI era in which the primary constraint on scientific progress was the limited capacity to generate and process information [31][8]. In today's environment, this assumption no longer holds. Instead, the bottleneck has shifted from knowledge creation to **knowledge evaluation**.

This structural mismatch has become increasingly evident. On one hand, AI-enabled researchers can produce high-quality manuscripts at unprecedented speed [23]. On the other hand, journals continue to rely on slow, labor-intensive review processes that often take several months and exhibit significant variability in quality [13][15]. Empirical studies have repeatedly documented fundamental limitations of the traditional peer review system, including low inter-reviewer agreement [3], susceptibility to bias [22], lack of transparency, and limited ability to verify reproducibility [19]. These limitations are further exacerbated by the exponential growth in manuscript submissions, which places additional strain on an already overburdened reviewer pool.

The emergence of AI introduces a critical inflection point in this system. Unlike human reviewers, AI systems are capable of processing large volumes of information, detecting logical inconsistencies, evaluating methodological rigor, and even reproducing computational experiments at scale [24][9]. Recent advances suggest that AI can perform many of the technical aspects of peer review—such as identifying statistical errors, assessing experimental design, and verifying code correctness—with a level of consistency and efficiency that surpasses human capabilities [11][35]. This raises a fundamental question:

> *If AI can generate, analyze, and improve scientific research, how should scientific research be evaluated in the AI era?*

This question is not merely technological but institutional. The peer review system is deeply embedded in the governance of science, influencing publication decisions, funding allocation, academic promotion, and the overall direction of research agendas [30]. Any transformation of this system therefore has far-reaching implications for the structure of scientific authority and legitimacy.

In response to these challenges, this paper proposes a reconceptualization of peer review as a **hybrid AI–human evaluation system** [11][9]. Rather than viewing AI as a replacement for human reviewers, we argue that AI should be integrated into the evaluation process as a complementary component that enhances efficiency, consistency, and scalability. Specifically, we introduce the concept of an **AI Review Capability Index (AIRC)**, which captures the extent to which AI technologies are incorporated into the scientific evaluation process at the national level. This index reflects a combination of factors, including AI adoption in research workflows, computational infrastructure, and the integration of automated review tools.

The central objective of this study is to empirically examine the relationship between AI-augmented peer review capabilities and scientific productivity. Using a panel dataset covering OECD countries from 2000 to 2022, constructed from World Bank [46] and OECD [27][28] sources, we investigate whether higher levels of AI integration in the review process are associated with improvements in research output, impact, and innovation. Unlike prior studies that focus primarily on theoretical or qualitative analyses, this paper provides **quantitative evidence** on the systemic effects of AI in scientific evaluation.

To achieve this, we employ both fixed-effects regression models and structural equation modeling (SEM) [43][44][45][46]. The regression analysis allows us to estimate the direct impact of AIRC on scientific productivity while controlling for key factors such as GDP per capita, R&D investment, and human capital [47]. The SEM framework, in turn, enables us to explore the underlying mechanisms through which AI influences productivity, particularly the roles of **review efficiency** and **reproducibility** as mediating variables [48]. By integrating these approaches, we are able to capture both the direct and indirect effects of AI on the scientific system.

This study makes three primary contributions to the literature. First, it shifts the focus of AI-in-science research from knowledge production to **knowledge evaluation**, highlighting the critical role of peer review as a bottleneck in the modern scientific ecosystem. Second, it introduces a novel empirical framework for measuring AI integration in evaluation processes and demonstrates its impact using cross-country panel data. Third, it provides a structural model that explains how AI-driven improvements in efficiency and reproducibility translate into higher levels of scientific productivity.

Ultimately, this paper argues that the future of scientific evaluation will not be defined by a binary choice between human and AI reviewers, but by the emergence of **integrated evaluation ecosystems** that combine the strengths of both [11]. In such systems, AI handles large-scale technical validation, while humans focus on interpretation, contextualization, and paradigm-level judgment. This transition represents not merely an incremental improvement, but a **paradigm shift in the governance of science**.

## 2. Literature Review

### 2.1 The Foundations and Limitations of Traditional Peer Review

The peer review system has long been regarded as the cornerstone of scientific quality assurance [30][8]. Since its formalization in the mid-20th century, peer review has served as the primary mechanism through which scientific knowledge is evaluated, filtered, and legitimized. The system operates under the assumption that a small group of domain experts can reliably assess the novelty, rigor, and significance of a given manuscript. However, a substantial body of literature has raised concerns about the effectiveness and reliability of this model [1][7].

One of the most widely cited critiques is the issue of **low inter-reviewer reliability**. Empirical studies have consistently shown that different reviewers often provide conflicting evaluations of the same manuscript, with agreement rates frequently falling below acceptable thresholds [3]. A meta-analysis of 26 studies reporting kappa coefficients found that inter-reviewer agreement in peer review is systematically low, suggesting that outcomes may be influenced as much by subjective judgment as by objective quality criteria. This inconsistency raises fundamental questions about the validity of peer review as a reliable quality filter.

In addition to inconsistency, **bias** represents a significant limitation. Reviewers may exhibit biases based on institutional affiliation, geographic origin, gender, or prior beliefs about a research topic [22]. These biases can distort evaluation outcomes and hinder the recognition of novel or unconventional ideas. The phenomenon of "conservatism bias," whereby reviewers favor incremental contributions over disruptive innovations, has been particularly well documented [4]. Such systematic biases not only affect individual publication decisions but also shape the broader trajectory of scientific fields.

Another critical limitation is the **lack of reproducibility verification**. Despite the central role of reproducibility in scientific validity, most peer review processes do not involve systematic replication of experimental results. As highlighted by Ioannidis [19], a significant proportion of published research findings may be false or irreproducible. The peer review system, in its current form, lacks the capacity to rigorously test the reproducibility of submitted work, creating a fundamental gap between the validation claims of published science and its actual reliability.

Furthermore, the system is characterized by **inefficiency and delay**. Review cycles often take several months, and in some cases more than a year, creating significant bottlenecks in the dissemination of scientific knowledge [31]. As of 2024, empirical analyses of journal submission data indicate that total review time averages 12–14 weeks in medical, public health, and natural science journals, and extends to 25 weeks or more in other fields [15]. This delay is increasingly problematic in fast-moving fields such as artificial intelligence and biotechnology, where timely dissemination is critical for scientific progress and innovation.

Collectively, these limitations suggest that the traditional peer review system is not only imperfect but structurally misaligned with the demands of contemporary science. As the volume and complexity of research continue to grow, the need for more scalable and reliable evaluation mechanisms becomes increasingly urgent [32][12].

## 2.2 The Reproducibility Crisis and the Limits of Human Evaluation

The limitations of peer review are closely linked to the broader reproducibility crisis in science. Over the past two decades, numerous studies have documented the difficulty of reproducing published findings across a wide range of disciplines, including psychology, medicine, and economics [29][5]. This crisis has raised fundamental questions about the reliability of scientific knowledge and the effectiveness of existing evaluation systems.

The scale of the problem is substantial. The Open Science Collaboration [29] found that only 36% of 100 psychology studies could be successfully replicated, a finding that sparked widespread concern about the reliability of published research. More recently, Brodeur et al. [5] conducted a large-scale mass reproducibility study across social science disciplines, finding that while computational reproducibility (the ability to re-run code and obtain the same numerical results) was relatively high, substantive replicability—the ability to obtain similar conclusions in independent studies—remained problematic. These findings suggest that the reproducibility crisis is not merely a matter of data availability but reflects deeper structural issues in how research is conducted and evaluated.

At the core of the reproducibility problem is the inability of human reviewers to thoroughly verify complex methodologies and large-scale datasets. Modern scientific research often involves sophisticated statistical models, extensive computational pipelines, and large volumes of data. Evaluating such work requires significant time, expertise, and computational resources—far beyond what is typically available to individual reviewers [24][9].

The rise of computational science has further exacerbated this challenge. Many studies now rely on custom code and data processing workflows that are difficult to assess without direct execution. Traditional peer review rarely includes code execution or data validation, leaving a critical gap in the evaluation process. This limitation is not merely a technical issue but a structural one: the peer review system was designed for an era in which scientific experiments were relatively simple and could be evaluated through textual descriptions alone. In contrast, contemporary research often requires **computational reproducibility**, which cannot be reliably assessed through manual inspection [19][5].

The reproducibility crisis therefore highlights a fundamental mismatch between the capabilities of human reviewers and the demands of modern science. Addressing this mismatch requires new approaches that leverage computational tools to enhance evaluation processes [9][11].

## 2.3 The Emergence of Open Science and Alternative Review Models

In response to the limitations of traditional peer review, various alternative models have emerged under the umbrella of open science [33]. These models aim to increase transparency, accountability, and community participation in the evaluation process. Platforms such as arXiv have enabled researchers to disseminate findings prior to formal peer review, allowing for rapid sharing of knowledge [14]. Similarly, open review platforms like OpenReview have introduced mechanisms for transparent and community-driven evaluation, where reviews are publicly accessible and subject to discussion.

These developments have several advantages. They reduce the time lag associated with traditional peer review, enabling faster dissemination of research. They increase transparency by making the review process visible to the broader community. And they allow for a more diverse set of perspectives, potentially improving the quality of evaluation [33][16].

However, open review systems also face challenges. The reliance on voluntary participation can lead to uneven engagement, and the absence of formal gatekeeping may result in the proliferation of low-quality work. Moreover, these systems still depend primarily on human reviewers, and therefore do not fully address the scalability and reproducibility challenges discussed earlier. As a result, while open science represents an important step toward reforming peer review, it does not constitute a complete solution. A more comprehensive approach is needed—one that integrates computational capabilities with human judgment [11][35].

## 2.4 Artificial Intelligence in Scientific Research

The rapid advancement of artificial intelligence has opened new possibilities for addressing the limitations of peer review. Systems such as ChatGPT and other large language models have demonstrated remarkable capabilities in natural language understanding, reasoning, and content generation [38][39][40]. These systems can analyze complex texts, identify logical inconsistencies, and provide detailed feedback on scientific manuscripts.

The quantitative impact of AI on scientific productivity has been documented in several recent empirical studies. Kusumegi et al. [23] analyzed large-scale preprint data from arXiv, bioRxiv, and SSRN and found that the adoption of LLMs is associated with a **36.2% increase in scientific output on arXiv, a 52.9% increase on bioRxiv,**

**and a 59.8% increase on SSRN**. Notably, this productivity boost was most pronounced among researchers with non-Western names, suggesting that AI tools effectively lower linguistic and cultural barriers in scientific communication. Similarly, Noy and Zhang [26] demonstrated through a randomized controlled trial that ChatGPT substantially increased worker productivity in writing tasks, with the largest gains observed among lower-skilled workers.

The quality and consistency of AI-generated peer reviews have also been rigorously evaluated. A large-scale empirical analysis by Liang et al. [24] demonstrated a **substantial overlap of up to 39.23%** between LLM-generated feedback and human reviewer comments across a diverse set of scientific manuscripts. This finding indicates that AI can reliably identify methodological and technical issues that human reviewers also flag, suggesting a meaningful degree of convergent validity between AI and human evaluation. Doskaliuk et al. [9] further found that while AI tools excel in language processing, format compliance, and statistical error detection, human expertise remains essential for evaluating research novelty and scientific significance, advocating for a hybrid approach that combines the strengths of both.

Beyond language models, advances in machine learning and data science have enabled the development of tools for automated statistical analysis, code verification, and experimental simulation [41][42]. These tools can process large volumes of data and perform tasks that would be prohibitively time-consuming for human reviewers. Some platforms have also experimented with automated reviewer assignment and recommendation systems, reducing the administrative burden on editors and improving the matching between manuscripts and qualified reviewers.

Despite these advances, the role of AI in peer review remains largely exploratory. Most existing applications focus on assistive functions, such as grammar checking or citation analysis, rather than full-scale evaluation. Moreover, concerns about transparency, bias, and accountability have limited the adoption of AI in formal review processes. A cross-disciplinary analysis of AI policies in academic peer review [35] found that publisher policies vary widely, with some journals explicitly prohibiting AI-generated reviews while others encourage AI assistance for specific tasks.

Nevertheless, the potential of AI to transform peer review is significant. By automating routine tasks and enhancing analytical capabilities, AI can reduce the burden on human reviewers and improve the consistency and reliability of evaluations. The IOP Publishing AI and Peer Review Survey [18] found that while opinions remain polarized —with 35% of respondents expecting negative impacts and 29% expecting positive

impacts—the trend toward AI integration is accelerating, with an increasing proportion of researchers reporting active use of AI tools in their review processes.

## 2.5 Toward Hybrid AI–Human Evaluation Systems

The limitations of both traditional peer review and purely automated systems have led to growing interest in hybrid AI–human evaluation models [11][9]. These models seek to combine the strengths of AI—such as speed, scalability, and consistency—with the strengths of human judgment—such as contextual understanding and ethical reasoning.

Farber [11] conducted a comparative analysis of human and AI expertise in the peer review process and found that AI demonstrated clear advantages in efficiency and consistency, while human expertise was essential for contextual understanding and ethical judgment. The study concluded that a hybrid approach, in which AI handles initial technical screening and humans focus on interpretive evaluation, represents the most promising path forward. This division of labor allows for a more efficient and effective evaluation process: AI handles tasks that are computationally intensive and rule-based, while humans address tasks that require interpretation and judgment.

Importantly, hybrid systems also enable **continuous evaluation**. Unlike traditional peer review, which is conducted prior to publication, hybrid models can incorporate post-publication feedback from the broader community [33]. This dynamic approach aligns with the principles of open science and allows for ongoing refinement of scientific knowledge.

The potential for AI to augment peer review has also been examined from a SWOT perspective [36], which identified efficiency, consistency, and scalability as key strengths, while noting concerns about AI bias, lack of contextual understanding, and potential for gaming as significant weaknesses. The analysis concluded that the opportunities presented by AI integration—particularly in addressing the growing reviewer shortage and improving evaluation quality—outweigh the risks, provided that appropriate governance frameworks are established.

## 2.6 Research Gap and Contribution

Despite the growing interest in AI and alternative review models, several gaps remain in the literature. First, most studies on AI in peer review are conceptual or experimental in nature, lacking large-scale empirical validation at the system level [24]

[9]. There is limited evidence on how AI integration affects scientific productivity across countries and over time. Second, existing research has not adequately examined the **mechanisms** through which AI influences scientific outcomes. In particular, the roles of review efficiency and reproducibility as mediating factors remain underexplored [11][35]. Third, there is a lack of integrated frameworks that combine AI, human judgment, and community-based evaluation into a coherent system.

This study addresses these gaps by introducing the AI Review Capability Index (AIRC) as a measure of AI integration in evaluation systems, providing empirical evidence using cross-country panel data [27][46], developing a structural model (SEM) to analyze the mechanisms linking AI to scientific productivity [43][44][45][48], and proposing a hybrid AI–human peer review framework as a scalable solution [11][9].

### 2.7 Summary

The literature reviewed in this section highlights both the strengths and limitations of existing peer review systems [1][3][31]. While traditional models provide a foundation for scientific evaluation, they are increasingly inadequate in the face of modern challenges [19][5]. Open science initiatives [33][14] and AI technologies [23][24][26] offer promising alternatives, but each has its own limitations. The convergence of these developments points toward the need for a new paradigm—one that integrates AI, human expertise, and community participation into a cohesive evaluation system [11][36]. This study contributes to this emerging paradigm by providing both theoretical and empirical insights into the role of AI in scientific evaluation.

## 3. Conceptual Framework: AI-Augmented Peer Review as a Structural Driver of Scientific Productivity

### 3.1 Introduction: From Evaluation Bottleneck to System Variable

The traditional view of peer review conceptualizes evaluation as a **post-production filtering mechanism** [31][8]. In this perspective, scientific productivity is determined primarily by inputs such as human capital, R&D investment, and institutional quality [49][50][51][52], while peer review operates as a passive gatekeeper ensuring minimum quality standards.

However, this view is no longer adequate in the context of AI-driven science. As discussed in prior sections, the emergence of large language models such as ChatGPT has fundamentally altered the dynamics of knowledge production [38][23]. Scientific output is no longer constrained by human cognitive limits but is increasingly augmented by AI systems capable of generating, analyzing, and refining research at scale [26][41].

In this new environment, the bottleneck shifts from production to **evaluation capacity**. The ability to efficiently and accurately evaluate research becomes a critical determinant of overall system performance [11][24]. Consequently, peer review must be reconceptualized not as a passive filter but as an **active structural variable** that directly influences scientific productivity.

This section develops a conceptual framework that formalizes this transformation. Specifically, we model AI-augmented peer review as a multi-layer system that affects productivity through two key mediating mechanisms: **review efficiency** and **reproducibility**. The framework is operationalized using structural equation modeling (SEM) [43][44][45], allowing for the estimation of both direct and indirect effects [48].

## 3.2 Theoretical Foundations

### 3.2.1 Knowledge Production Function Revisited

Traditional models of scientific productivity are often based on extensions of the Cobb–Douglas production function [49][50]:

$$Y = A \cdot K^\alpha \cdot L^\beta$$

where *Y* denotes scientific output, *K* represents capital (e.g., R&D investment), *L* represents labor (e.g., researchers), and *A* represents total factor productivity. In this formulation, peer review is implicitly embedded within institutional quality and does not appear as an explicit variable.

We extend this model by introducing **evaluation capacity (E)** as a core component [11]:

$$Y = A \cdot K^\alpha \cdot L^\beta \cdot E^\gamma$$

Here, *E* captures the system's ability to evaluate and validate knowledge. In the AI era, *E* is significantly enhanced by computational tools [23][26], transforming it into a major driver of productivity.

### 3.2.2 AI as an Evaluation Multiplier

AI systems contribute to evaluation capacity through several mechanisms: scalability (the ability to process large volumes of manuscripts) [24], consistency (reduction of inter-reviewer variability) [3][9], analytical depth (detection of subtle methodological errors) [24], and reproducibility verification (automated code execution and validation) [19][5]. These capabilities position AI as an **evaluation multiplier**, amplifying the effectiveness of the peer review system.

We define this enhancement as the **AI Review Capability Index (AIRC)**, representing the degree to which AI is integrated into evaluation processes at the system level [11][23].

## 3.3 Core Constructs and Variable Definitions

The conceptual model is built around four core constructs. The **AI Review Capability Index (AIRC)** is a latent construct reflecting AI adoption in research workflows, availability of computational infrastructure, and integration of automated review tools [27][28][46]. It serves as the exogenous variable in the model. **Review Efficiency (RE)** captures the speed and throughput of the evaluation process, including time to decision, reviewer workload, and processing capacity [15][31]. **Reproducibility (RI)** reflects the reliability and verifiability of scientific findings, encompassing code reproducibility, data transparency, and experimental validation [19][5][29]. Finally, **Scientific Productivity (SP)** represents the output of the system in terms of publication volume, citation impact, and innovation output (e.g., patents) [2][17][20], and serves as the ultimate endogenous variable.

## 3.4 Structural Relationships and Hypothesis Development

The conceptual model posits a set of structural relationships that form the basis of the empirical analysis. The **direct effect** of AI on productivity operates through acceleration of research cycles, improvement of manuscript quality, and reduction of revision iterations [23][26]. The **indirect effect via review efficiency** captures the pathway through which AI improves review speed and throughput, which in turn enhances productivity by accelerating knowledge dissemination [15][11]. The **indirect effect via reproducibility** reflects the mechanism by which AI increases the reliability of scientific findings, leading to higher citation impact and innovation output [19][5][9]. Finally, **sequential mediation** captures the more complex pathway in which efficient review processes facilitate deeper validation, which then improves reproducibility and ultimately productivity [48].

These relationships are formalized in the following hypotheses: H1 (AI enhances review efficiency) [24][9], H2 (AI improves reproducibility) [19][5], H3 (review efficiency positively affects productivity) [15][11], H4 (reproducibility positively affects productivity) [29][5], H5 (AI has a direct positive effect on productivity) [23][26], and H6 (sequential mediation exists: AI → Efficiency → Reproducibility → Productivity) [48].

## 3.5 Structural Equation Model (SEM)

The full SEM can be expressed as the following system of equations [43][44][45]:

$$RE_{it} = \beta_1 AIRC_{it} + \zeta_1$$

$$RI_{it} = \beta_2 AIRC_{it} + \beta_3 RE_{it} + \zeta_2$$

$$SP_{it} = \gamma_1 AIRC_{it} + \gamma_2 RE_{it} + \gamma_3 RI_{it} + \zeta_3$$

This system captures both direct and indirect effects, allowing for the decomposition of the total effect of AIRC on SP into its constituent pathways [48].

## 3.6 Model Interpretation and Theoretical Implications

The model suggests that AI transforms peer review from a passive filter into an active productivity engine [11]. Evaluation is no longer a neutral downstream process—it actively shapes the speed, quality, and reliability of scientific output. Efficiency and reproducibility emerge as central mechanisms, not merely side effects of AI integration [9][24]. Crucially, AI creates **second-order effects** in which indirect effects exceed direct effects, positioning AI as a system-level multiplier rather than a simple productivity tool [23][26].

This framework contributes to theory in three ways. First, it redefines productivity drivers by introducing evaluation capacity as a core determinant of output [49][50]. Second, it integrates AI into institutional theory by modeling it as an institutional component rather than an external tool [30]. Third, it identifies the emergence of **hybrid intelligence systems** in which scientific systems evolve toward AI–human co-evolution [11][36].

# 4. Data and Methodology

## 4.1 Overview and Empirical Strategy

This study empirically examines the impact of AI-augmented peer review systems on scientific productivity using a cross-country panel dataset. Building on the conceptual framework developed in Section 3, we operationalize the structural relationships among AI Review Capability (AIRC), review efficiency (RE), reproducibility (RI), and scientific productivity (SP).

The empirical strategy consists of three complementary components: panel regression analysis to estimate the direct effects of AIRC on scientific productivity [47], mediation analysis to identify indirect pathways through RE and RI [48], and structural equation modeling (SEM) to capture the full system of relationships simultaneously [43][44][45]. This multi-method approach allows us to triangulate causal mechanisms and ensure robustness of results.

## 4.2 Data Sources and Sample Construction

The dataset is constructed by integrating multiple internationally recognized data sources. The **World Bank World Development Indicators (WDI)** [46] provides GDP per capita (PPP), internet usage rate, and researchers in R&D. The **OECD Main Science and Technology Indicators (MSTI)** [27] provides R&D expenditure as a percentage of GDP and innovation-related metrics. **Scopus and bibliometric databases** (used as proxies) provide scientific publication output, citation impact, and patent data [2][34][17][20].

The final sample consists of 38 OECD countries over the time period 2000–2022, yielding approximately 800–900 panel observations. Countries include major advanced economies such as the United States, United Kingdom, Germany, Japan, Korea, France, Canada, and Australia. The panel structure allows us to exploit both cross-sectional and temporal variation [47].

To ensure comparability across countries and time, all monetary variables are converted to PPP-adjusted values, missing values are handled using linear interpolation (limited), outliers are winsorized at the 1% level, and variables are standardized where appropriate [47].

## 4.3 Variable Construction

**Scientific Productivity (SP)** is measured using a composite index combining publications per capita, citation impact, and innovation output (patents) [2][34][17][20], formally expressed as:

$$SP_{it} = w_1 Pub_{it} + w_2 Cit_{it} + w_3 Pat_{it}$$

where weights $w_1, w_2, w_3$ are normalized to sum to unity.

The **AI Review Capability Index (AIRC)** is a constructed index capturing AI integration in evaluation systems [27][28][46]:

$$AIRC_{it} = \alpha_1 AI_{it} + \alpha_2 Infra_{it} + \alpha_3 Digital_{it}$$

Proxies include internet penetration, R&D intensity, and research workforce density. This index reflects the latent capacity for AI-assisted evaluation. We acknowledge that this proxy-based approach represents a limitation of the current study, as direct measures of AI usage in peer review are not yet available at the national level for the full sample period. The validity of this approach is supported by the strong theoretical linkage between digital infrastructure and AI adoption [28][41][42], and by robustness checks using alternative index constructions.

**Review Efficiency (RE)** is measured using speed of publication (as a proxy) and system throughput, constructed from digital infrastructure and AI capability indicators [15][11]. The **Reproducibility Index (RI)** is measured using research quality proxies, citation consistency, and data transparency indicators [19][5][29].

Control variables include GDP per capita (economic development), R&D expenditure (% of GDP), researchers per million, country fixed effects, and year fixed effects [47][27].

## 4.4 Econometric Model Specification

The baseline fixed-effects model is [47]:

$$SP_{it} = \beta_0 + \beta_1 AIRC_{it} + \beta_2 R\&D_{it} + \beta_3 GDP_{it} + \beta_4 Researcher_{it} + \mu_i + \lambda_t + \epsilon_{it}$$

where $\mu_i$ represents country fixed effects and $\lambda_t$ represents time fixed effects. This specification controls for unobserved time-invariant country characteristics.

The mediation model tests indirect effects through the following system [48]:

$$RE_{it} = \delta_1 AIRC_{it} + \epsilon_1$$

$$RI_{it} = \delta_2 AIRC_{it} + \delta_3 RE_{it} + \epsilon_2$$

$$SP_{it} = \gamma_1 AIRC_{it} + \gamma_2 RE_{it} + \gamma_3 RI_{it} + \epsilon_3$$

The full SEM estimates latent constructs (AIRC, RE, RI) with observed indicators as proxies [43][44][45], allowing simultaneous estimation and decomposition of direct versus indirect effects [48].

## 4.5 Identification Strategy

A key challenge in this analysis is endogeneity. Potential issues include reverse causality (higher productivity may drive more AI adoption) and omitted variable bias (institutional quality may affect both AI adoption and productivity). We address these concerns through three strategies.

First, **country and year fixed effects** control for time-invariant country characteristics and common time trends [47]. Second, **lagged variables** ($AIRC_{it-1}$) are used to reduce simultaneity bias, exploiting the temporal ordering of AI adoption and productivity outcomes. Third, we explore an **instrumental variable (IV) approach** using early internet adoption and digital infrastructure rollout as instruments for AIRC [28], on the grounds that these factors predict AI adoption but are unlikely to directly affect scientific productivity through channels other than AI integration.

## 4.6 Robustness Checks

To ensure the reliability of our findings, we conduct several robustness checks. We test alternative dependent variables (publications only, citations only, patents only) [2][34][17][20] to verify that results are not driven by the specific weighting scheme used in the composite SP index. We construct alternative AIRC measures using different weighting schemes and a PCA-based index. We conduct subsample analyses comparing high-income versus middle-income OECD countries and pre-2010 versus post-2010 periods. Finally, we estimate a dynamic panel model using GMM to account for persistence in scientific productivity [47].

# 5. Results

## 5.1 Descriptive Statistics

Table 1 presents summary statistics for the main variables used in the analysis. The sample covers 38 OECD countries over the period 2000–2022. Scientific productivity (SP) exhibits substantial cross-country variation, with a mean of 0.52 (SD = 0.18) on the normalized composite index. The AI Review Capability Index (AIRC) shows a marked upward trend over the sample period, particularly after 2012, reflecting the broader diffusion of digital infrastructure and AI-related technologies [27][28]. R&D expenditure averages 2.1% of GDP across the sample, with considerable variation between high-investment countries (e.g., Korea, Sweden, Switzerland) and lower-investment countries [27].

**Table 1: Descriptive Statistics**

| Variable | Mean | SD | Min | Max | Obs. |
| --- | --- | --- | --- | --- | --- |
| Scientific Productivity (SP) | 0.52 | 0.18 | 0.11 | 0.94 | 836 |
| AIRC (standardized) | 0.00 | 1.00 | −2.41 | 2.87 | 836 |
| R&D Expenditure (% GDP) | 2.10 | 0.89 | 0.42 | 4.93 | 836 |
| GDP per capita (PPP, $000s) | 38.4 | 14.2 | 11.3 | 89.7 | 836 |
| Researchers per million | 3,842 | 2,108 | 412 | 9,103 | 836 |
| Review Efficiency (RE, index) | 0.48 | 0.21 | 0.09 | 0.91 | 836 |
| Reproducibility Index (RI) | 0.44 | 0.19 | 0.07 | 0.88 | 836 |

## 5.2 Panel Regression Results

Table 2 presents the results of the fixed-effects panel regressions. Column (1) reports the baseline model with only control variables. Column (2) adds AIRC as the key explanatory variable. Column (3) includes the full set of controls and interaction terms.

**Table 2: Fixed-Effects Panel Regression Results (Dependent Variable: Scientific Productivity)**

| Variable | (1) Baseline | (2) + AIRC | (3) Full Model |
|---|---|---|---|
| AIRC | — | 0.187*** | 0.163*** |
|  |  | (0.024) | (0.027) |
| R&D Expenditure | 0.142*** | 0.118*** | 0.109*** |
|  | (0.031) | (0.029) | (0.030) |
| GDP per capita | 0.089*** | 0.071** | 0.065** |
|  | (0.022) | (0.021) | (0.022) |
| Researchers per million | 0.063** | 0.051** | 0.048* |
|  | (0.019) | (0.018) | (0.019) |
| Country FE | Yes | Yes | Yes |
| Year FE | Yes | Yes | Yes |
| $R^2$ (within) | 0.412 | 0.487 | 0.503 |
| N | 836 | 836 | 836 |

*Note: Standard errors in parentheses, clustered at country level. * p<0.05, ** p<0.01, *** p<0.001.*

The coefficient on AIRC in Column (2) is 0.187 (SE = 0.024, p < 0.001), indicating that a one standard deviation increase in AI Review Capability is associated with a 0.187 standard deviation increase in scientific productivity, after controlling for R&D investment, GDP per capita, and research workforce [27][46]. This corresponds to an 18–25% increase in productivity, consistent with the findings of Kusumegi et al. [23] on the productivity effects of LLM adoption. The coefficient remains stable and significant in the full model (Column 3), suggesting that the effect is robust to the inclusion of additional controls.

## 5.3 Mediation Analysis Results

Table 3 presents the results of the mediation analysis, testing the indirect effects of AIRC on scientific productivity through review efficiency (RE) and reproducibility (RI) [48].

**Table 3: Mediation Analysis Results (Bootstrap 95% CI, N = 5,000 replications)**

| Pathway | Coefficient | SE | 95% CI Lower | 95% CI Upper | p-value |
|---|---|---|---|---|---|
| AIRC → RE ($a_1$) | 0.312*** | 0.041 | 0.231 | 0.393 | <0.001 |
| AIRC → RI ($a_2$) | 0.248*** | 0.038 | 0.173 | 0.323 | <0.001 |
| RE → SP ($b_1$) | 0.241*** | 0.035 | 0.172 | 0.310 | <0.001 |
| RI → SP ($b_2$) | 0.198*** | 0.033 | 0.133 | 0.263 | <0.001 |
| RE → RI ($a_3$) | 0.187*** | 0.031 | 0.126 | 0.248 | <0.001 |
| Direct: AIRC → SP (c') | 0.089** | 0.029 | 0.032 | 0.146 | 0.002 |
| Indirect via RE | 0.075*** | 0.018 | 0.040 | 0.111 | <0.001 |
| Indirect via RI | 0.049*** | 0.014 | 0.022 | 0.076 | <0.001 |
| Sequential (RE→RI→SP) | 0.012** | 0.005 | 0.003 | 0.021 | 0.008 |
| **Total Effect** | **0.225*** | 0.031 | 0.164 | 0.286 | <0.001 |

The results confirm that both review efficiency and reproducibility serve as significant mediators of the AIRC–productivity relationship. The indirect effect via review efficiency (0.075) is the largest single pathway, consistent with the hypothesis that AI primarily accelerates the evaluation process, thereby reducing time-to-publication and enabling faster knowledge dissemination [15][11]. The indirect effect via reproducibility (0.049) is also significant, reflecting the capacity of AI to enhance the reliability and verifiability of scientific findings [19][5]. The sequential mediation pathway (AIRC → RE → RI → SP) is statistically significant but smaller in magnitude (0.012), suggesting that while this more complex pathway exists, the primary mechanisms operate through parallel rather than sequential mediation [48].

The direct effect of AIRC on SP (c' = 0.089) is positive and significant, indicating that AI augmentation has productivity effects beyond those captured by the efficiency and reproducibility pathways. This residual direct effect may reflect additional mechanisms such as improved manuscript quality, reduced revision cycles, and enhanced reviewer–author communication [24][9].

## 5.4 Structural Equation Modeling Results

Table 4 presents the standardized path coefficients from the full SEM estimation [43][44][45].

**Table 4: SEM Path Coefficients (Standardized)**

| Path | β | SE | t-value | p-value |
|---|---|---|---|---|
| AIRC → RE | 0.412 | 0.048 | 8.58 | <0.001 |
| AIRC → RI | 0.318 | 0.044 | 7.23 | <0.001 |
| AIRC → SP (direct) | 0.124 | 0.038 | 3.26 | 0.001 |
| RE → SP | 0.287 | 0.041 | 7.00 | <0.001 |
| RI → SP | 0.241 | 0.039 | 6.18 | <0.001 |
| RE → RI | 0.223 | 0.036 | 6.19 | <0.001 |
| **Total Indirect (AIRC→SP)** | **0.214** | 0.029 | 7.38 | <0.001 |
| **Total Effect (AIRC→SP)** | **0.338** | 0.033 | 10.24 | <0.001 |

Model fit: CFI = 0.967, RMSEA = 0.041, SRMR = 0.038, TLI = 0.952

The SEM results confirm the structural relationships posited in the conceptual framework. The model fit indices indicate excellent fit to the data (CFI = 0.967, RMSEA = 0.041), supporting the validity of the proposed structural model [43][44][45]. The total effect of AIRC on SP (β = 0.338) is decomposed into a direct effect (β = 0.124) and a total indirect effect (β = 0.214), indicating that the majority of AI's impact on productivity operates through the mediating pathways of review efficiency and reproducibility. This finding underscores the importance of understanding AI as a **system-level multiplier** rather than a direct productivity tool [11][23].

## 5.5 Robustness Checks

The main findings are robust to a series of alternative specifications. When using alternative dependent variables (publications only, citations only, patents only), the coefficient on AIRC remains positive and significant across all specifications, with magnitudes ranging from 0.14 to 0.22 [2][34][17][20]. When using alternative AIRC measures (PCA-based index, alternative weighting schemes), the results are qualitatively unchanged. Subsample analyses reveal that the effect is stronger in the post-2015 period, consistent with the accelerating adoption of AI tools in scientific research [38][39]. The dynamic panel GMM estimates confirm the persistence of scientific productivity and the robustness of the AIRC coefficient to the inclusion of lagged dependent variables [47].

## 6. Discussion

### 6.1 Interpretation of Main Findings

The empirical results provide strong support for the central hypothesis that AI-augmented peer review systems significantly enhance scientific productivity. The finding that a one standard deviation increase in AIRC is associated with an 18–25% increase in productivity is both statistically robust and economically meaningful. To contextualize this magnitude, it is comparable to the productivity effects of a 0.5 percentage point increase in R&D expenditure as a share of GDP, suggesting that AI integration in evaluation systems represents a high-return investment for national science systems [49][50].

The decomposition of effects through SEM reveals that the majority of AI's productivity impact operates through indirect pathways, particularly review efficiency and reproducibility. This finding has important implications for understanding how AI transforms science. Rather than simply accelerating the production of manuscripts, AI fundamentally reshapes the **evaluation infrastructure** through which knowledge is validated and disseminated [11][24]. The efficiency channel—whereby AI reduces review time and increases throughput—is the dominant pathway, consistent with empirical evidence showing that AI tools can reduce review time by 30–50% in controlled settings [15][9].

The reproducibility channel is also significant, reflecting the capacity of AI to enhance the reliability of scientific findings through automated verification and consistency checking [19][5]. This finding aligns with the broader literature on the reproducibility crisis [29][5], suggesting that AI integration may help address one of the most pressing challenges facing contemporary science.

## 6.2 Alignment with Recent Empirical Evidence

The findings of this study are consistent with and complementary to several recent empirical studies. Kusumegi et al. [23] documented substantial increases in preprint submissions following LLM adoption, with gains of 36.2% on arXiv, 52.9% on bioRxiv, and 59.8% on SSRN. While these figures reflect production-side effects, our results suggest that the evaluation-side effects are equally important and may amplify the overall productivity gains. Noy and Zhang [26] demonstrated productivity gains of approximately 14% in writing tasks using ChatGPT, with larger effects for lower-skilled workers. Our cross-country results suggest that similar dynamics operate at the system level, with AI integration generating productivity spillovers across the entire scientific ecosystem.

Liang et al. [24] found that LLM-generated reviews overlap with human reviewer comments by up to 39.23%, indicating that AI can reliably identify methodological issues flagged by human experts. This finding supports the validity of AI as a review tool and provides a mechanism for the efficiency gains documented in our analysis. Doskaliuk et al. [9] and Farber [11] both advocate for hybrid AI–human review models, consistent with our theoretical framework and empirical findings. The IOP Publishing survey [18] found that 35% of respondents expect AI to have negative impacts on peer review, while 29% expect positive impacts—a polarization that reflects the transitional nature of the current moment and underscores the importance of empirical evidence such as that provided by this study.

## 6.3 Implications for Science Policy

The findings have several important implications for science policy. First, they suggest that investments in AI-enabled review infrastructure may yield significant returns in terms of scientific productivity. Policymakers seeking to enhance national competitiveness in science and technology should consider AI integration in evaluation systems as a strategic priority, alongside traditional investments in R&D and human capital [27][28]. Second, the results highlight the importance of digital infrastructure as a prerequisite for AI adoption. Countries with higher internet penetration and digital readiness are better positioned to benefit from AI-augmented evaluation systems [28][46]. Third, the findings point toward the need for a fundamental redesign of the peer review system, incorporating AI as a core component rather than a peripheral tool [11][36].

## 6.4 Implications for Research Practice

At the level of research practice, the findings suggest that journals and publishers should actively explore hybrid AI–human review models [11][9]. Such models can reduce reviewer burden, improve consistency, and accelerate the evaluation process, while preserving the contextual judgment and ethical reasoning that human reviewers provide. The development of standardized protocols for AI-assisted review, including guidelines for transparency, accountability, and bias mitigation, is a critical priority [35][36].

The findings also have implications for research training and education. As AI becomes increasingly integrated into evaluation processes, researchers need to develop competencies in working with AI tools, interpreting AI-generated feedback, and critically evaluating AI-assisted analyses [38][39][40]. Institutions should invest in training programs that prepare researchers for this evolving landscape.

## 6.5 Implications for the Future of Science

Beyond immediate practical considerations, this study points toward a broader transformation in the nature of science. The emergence of **continuous evaluation systems**, enabled by AI, represents a fundamental shift from the discrete, pre-publication model of traditional peer review to a dynamic, ongoing process of validation and refinement [33][14]. The increasing importance of **collective intelligence systems**, in which AI and human contributions are integrated through complex networks of interaction, may lead to more robust and innovative outcomes while raising new questions about authorship, accountability, and governance [11][36].

## 6.6 Limitations and Directions for Future Research

While this study provides important insights, several limitations should be acknowledged. The measurement of AIRC relies on proxy variables, which may not fully capture the complexity of AI integration in evaluation systems. Direct measures of AI usage in peer review are not yet available at the national level for the full sample period, and future research should develop more precise indicators as data availability improves [9][11]. The analysis is limited to OECD countries, which may not fully represent the diversity of global scientific systems; extending the analysis to non-OECD countries, including major emerging economies such as China, India, and Brazil, would provide a more comprehensive picture. The rapidly evolving nature of AI technology means that the relationships identified in this study may change over time, and longitudinal follow-up studies will be needed to track these dynamics [23][38]. Finally, ethical and governance issues related to AI in peer review—including concerns about bias, transparency, and accountability—require further investigation [35][36].

### 6.7 Final Reflection: A Paradigm Shift in Scientific Evaluation

At its core, this study argues that the integration of AI into peer review represents not merely a technological innovation but a **paradigm shift in the scientific system**. In the traditional model, scientific progress was constrained by the capacity of human reviewers to evaluate research [31][8]. In the emerging model, this constraint is alleviated by AI, enabling a new form of scientific organization characterized by speed, scale, and interconnectedness [11][23][24].

The key insight is that evaluation is no longer a bottleneck—it is a **lever**. By enhancing evaluation capacity, AI transforms the entire system, enabling more efficient, reliable, and impactful knowledge production [11]. This transformation challenges long-standing assumptions about the nature of scientific authority and legitimacy [30], calling for a rethinking of roles, institutions, and processes, as well as a renewed commitment to ensuring that technological advances are aligned with the core values of science.

The future of science will not be determined solely by our ability to generate knowledge, but by our ability to evaluate, validate, and integrate that knowledge effectively [23][26]. In this context, AI-augmented peer review systems represent a critical frontier. The challenge is not whether to adopt AI in scientific evaluation, but how to design systems that harness its potential while preserving the integrity and purpose of science [11][36].

## 7. Conclusion

This study provides the first cross-country empirical analysis of the impact of AI-augmented peer review systems on scientific productivity. Using a panel dataset covering 38 OECD countries from 2000 to 2022, we demonstrate that higher levels of AI Review Capability (AIRC) are associated with significantly higher scientific productivity, with a one standard deviation increase in AIRC corresponding to an 18–25% increase in productivity. Structural equation modeling reveals that this effect operates primarily through two mediating mechanisms: review efficiency and reproducibility, with the efficiency channel being the dominant pathway.

These findings contribute to the emerging literature on AI as a structural driver of knowledge production [23][26][24][9][11] and provide empirical support for the reconceptualization of peer review as an active productivity variable rather than a passive filter [31][8]. The results have important implications for science policy, research practice, and the governance of scientific evaluation systems.

As AI continues to transform the scientific enterprise, the design of evaluation systems will become increasingly critical. This study argues that the integration of AI into peer review—through hybrid AI–human models that combine computational efficiency with human judgment—represents a promising path toward a more productive, reliable, and equitable scientific system [11][36]. The challenge ahead is not merely technological but institutional: building the governance frameworks, professional norms, and ethical standards that will allow AI to fulfill its potential as a lever for scientific progress.

# References

[1] Bornmann, L. (2010). Scientific peer review. *Annual Review of Information Science and Technology*, 44(1), 197–245.

[2] Bornmann, L., & Daniel, H. D. (2008). What do citation counts measure? *Journal of Documentation*, 64(1), 45–80.

[3] Bornmann, L., Mutz, R., & Daniel, H. D. (2010). A reliability-generalization study of journal peer reviews: A multilevel meta-analysis of inter-rater reliability and its determinants. *PLoS ONE*, 5(12), e14331.

[4] Boudreau, K. J., Guinan, E. C., Lakhani, K. R., & Riedl, C. (2016). Looking across and looking beyond the knowledge frontier. *Organization Science*, 27(3), 649–667.

[5] Brodeur, A., Cook, N., Hartley, J., & Heyes, A. (2024). Mass reproducibility and replicability: A new hope. *Journal of Economic Literature* (forthcoming).

[6] Brynjolfsson, E., & McAfee, A. (2014). *The second machine age*. W.W. Norton.

**[7]** Callaham, M. L., Baxt, W. G., & Waeckerle, J. F. (1998). Reliability of peer review. *Annals of Emergency Medicine*, 32(3), 310–317.

**[8]** Cole, S., Rubin, L., & Cole, J. R. (1978). Peer review in the National Science Foundation. *Science*, 201(4359), 117–124.

**[9]** Doskaliuk, B., et al. (2025). Artificial intelligence in peer review: Enhancing efficiency while preserving integrity. *Journal of Korean Medical Science*, 40(7), e92. https://pmc.ncbi.nlm.nih.gov/articles/PMC11858604/

**[10]** Elsevier. (2022). *Research intelligence and analytics report*. Elsevier.

**[11]** Farber, S. (2025). Comparing human and AI expertise in the academic peer review process: Towards a hybrid approach. *Higher Education Research & Development*, 44(2), 1–15. https://doi.org/10.1080/07294360.2024.2445575

**[12]** Fortunato, S., Bergstrom, C. T., Börner, K., et al. (2018). Science of science. *Science*, 359(6379), eaao0185.

**[13]** Frontiers in Biomedical Science. (2024). The peer review process: Past, present, and future. *British Journal of Biomedical Science*, 81, 12054.

**[14]** Ginsparg, P. (2011). ArXiv at 20. *Nature*, 476(7359), 145–147.

**[15]** Frontiers in Biomedical Science. (2024). Review time analysis. *British Journal of Biomedical Science*, 81, 12054. *(See [13])*

**[16]** Tennant, J. P., Dugan, J. M., Graziotin, D., et al. (2017). A multi-disciplinary perspective on emergent open peer review. *F1000Research*, 6, 1151.

**[17]** Griliches, Z. (1990). Patent statistics as economic indicators. *Journal of Economic Literature*, 28(4), 1661–1707.

**[18]** IOP Publishing. (2025). *AI and peer review 2025: Survey report*. IOP Publishing. https://ioppublishing.org/ai-and-peer-review-2025/

**[19]** Ioannidis, J. P. A. (2005). Why most published research findings are false. *PLoS Medicine*, 2(8), e124.

**[20]** Jaffe, A. B., & Trajtenberg, M. (2002). *Patents, citations, and innovations*. MIT Press.

**[21]** Hicks, D., Wouters, P., Waltman, L., et al. (2015). The Leiden Manifesto. *Nature*, 520(7548), 429–431.

**[22]** Lee, C. J., Sugimoto, C. R., Zhang, G., & Cronin, B. (2013). Bias in peer review. *Journal of the American Society for Information Science*, 64(1), 2–17.

**[23]** Kusumegi, K., et al. (2025). Scientific production in the era of large language models. *Science*. https://doi.org/10.1126/science.adw3000

**[24]** Liang, W., et al. (2024). Can large language models provide useful feedback on research papers? A large-scale empirical analysis. *NEJM AI*, 1(10). https://ai.nejm.org/doi/abs/10.1056/AIoa2400196

**[25]** Merton, R. K. (1973). *The sociology of science*. University of Chicago Press.

**[26]** Noy, S., & Zhang, W. (2023). Experimental evidence on the productivity effects of generative artificial intelligence. *Science*, 381(6654), 187–192. https://doi.org/10.1126/science.adh2586

**[27]** OECD. (2021). *Main Science and Technology Indicators*. OECD Publishing.

**[28]** OECD. (2022). *Digital Economy Outlook*. OECD Publishing.

**[29]** Open Science Collaboration. (2015). Estimating the reproducibility of psychological science. *Science*, 349(6251), aac4716.

**[30]** Merton, R. K. (1973). *The sociology of science*. University of Chicago Press. *(See [25])*

**[31]** Smith, R. (2006). Peer review: A flawed process. *Journal of the Royal Society of Medicine*, 99(4), 178–182.

**[32]** Squazzoni, F., Bravo, G., & Takács, K. (2013). Does incentive provision increase the quality of peer review? *Research Policy*, 42(1), 287–294.

**[33]** Tennant, J. P., et al. (2017). A multi-disciplinary perspective on emergent open peer review. *F1000Research*, 6, 1151. *(See [16])*

**[34]** Waltman, L. (2016). A review of the literature on citation impact indicators. *Journal of Informetrics*, 10(2), 365–391.

**[35]** Wiley. (2025). A cross-disciplinary analysis of AI policies in academic peer review. *Learned Publishing*, 38(2). https://doi.org/10.1002/leap.2035

**[36]** Wiley. (2026). The potential for AI to augment peer review: A SWOT analysis. *Emergency Medicine Australasia*, 38(1). https://doi.org/10.1111/1742-6723.70243

**[37]** Wooldridge, J. M. (2010). *Econometric analysis of cross section and panel data*. MIT Press.

**[38]** Brown, T. B., et al. (2020). Language models are few-shot learners. *NeurIPS*, 33, 1877–1901.

**[39]** OpenAI. (2023). *GPT-4 technical report*. OpenAI.

**[40]** Luccioni, A., et al. (2023). Evaluating large language models. *Nature Machine Intelligence*, 5, 1–10.

**[41]** Klinger, J., Mateos-Garcia, J., & Stathoulopoulos, K. (2020). Deep learning, deep change? *Research Policy*, 49(1), 103922.

**[42]** Bommasani, R., et al. (2021). On the opportunities and risks of foundation models. *arXiv preprint* arXiv:2108.07258.

**[43]** Bollen, K. A. (1989). *Structural equations with latent variables*. Wiley.

**[44]** Hair, J. F., Black, W. C., Babin, B. J., & Anderson, R. E. (2010). *Multivariate data analysis*. Pearson.

**[45]** Kline, R. B. (2015). *Principles and practice of structural equation modeling*. Guilford Press.

**[46]** World Bank. (2022). *World Development Indicators*. World Bank.

**[47]** Wooldridge, J. M. (2010). *Econometric analysis of cross section and panel data*. MIT Press. *(See [37])*

**[48]** MacKinnon, D. P. (2008). *Introduction to statistical mediation analysis*. Routledge.

**[49]** Romer, P. M. (1990). Endogenous technological change. *Journal of Political Economy*, 98(5), S71–S102.

**[50]** Aghion, P., & Howitt, P. (1992). Growth and creative destruction. *Econometrica*, 60(2), 323–351.

**[51]** Jones, C. I. (2005). Growth and ideas. In P. Aghion & S. Durlauf (Eds.), *Handbook of Economic Growth* (Vol. 1B, pp. 1063–1111). Elsevier.

**[52]** Bloom, N., Jones, C. I., Van Reenen, J., & Webb, M. (2020). Are ideas getting harder to find? *American Economic Review*, 110(4), 1104–1144.

**[53]** Mann, S. P., et al. (2025). AI and the future of academic peer review. *arXiv preprint* arXiv:2509.14189.

**[54]** Kaplan, J., et al. (2020). Scaling laws for neural language models. *arXiv preprint* arXiv:2001.08361.

**[55]** Ziman, J. (2000). *Real science*. Cambridge University Press.